# Accurate dynamical mass determination of a classical Cepheid in an eclipsing binary system

G. Pietrzyński[1,2], I.B. Thompson[3], W. Gieren[1], D. Graczyk[1], G. Bono[4,5], A. Udalski[2], I. Soszyński[2], D. Minniti[6], B. Pilecki[1,2]

1. Universidad de Concepción, Departamento de Astronomìa, Casilla 160-C, Concepciòn, Chile

2. Obserwatorium Astronomiczne Uniwersytetu Warszawskiego, Aleje Ujazdowskie 4, 00-478 Warszawa, Poland

3. Carnegie Observatories, 813 Santa Barbara Street, Pasadena, CA 911101-1292, USA

4. Dipartimento di Fisica Universita' di Roma Tor Vergata, via della Ricerca Scientifica 1, 00133 Rome, Italy

5. INAF-Osservatorio Astronomico di Roma, Via Frascati 33, 00040 Monte Porzio Catone, Italy

6. Pontificia Universidad Católica de Chile, Departamento de Astronomía y Astrofísica, Casilla 306, Santiago 22, Chile

**Stellar pulsation theory provides a means of determining the masses of pulsating classical Cepheid supergiants—it is the pulsation that causes their luminosity to vary. Such pulsational masses are found to be smaller than the masses derived from stellar evolution theory: this is the Cepheid mass discrepancy problem (1,2), for which a solution is missing (3–5). An independent, accurate dynamical mass determination for a classical Cepheid variable star (as opposed to type-II Cepheids, low-mass stars with a very different evolutionary history) in a binary system is needed in order to determine which is correct. The accuracy of previous efforts to establish a dynamical Cepheid mass from Galactic single-lined non-eclipsing binaries was typically about 15–30 per cent (refs 6, 7), which is not good**

**enough to resolve the mass discrepancy problem. In spite of many observational efforts (8,9), no firm detection of a classical Cepheid in an eclipsing double-lined binary has hitherto been reported. Here we report the discovery of a classical Cepheid in a well detached, double-lined eclipsing binary in the Large Magellanic Cloud. We determine the mass to a precision of one per cent and show that it agrees with its pulsation mass, providing strong evidence that pulsation theory correctly and precisely predicts the masses of classical Cepheids.**

In the course of the OGLE microlensing survey conducted by several members of our We have detected several candidates for Cepheid variables in eclipsing binary systems in the Large Magellanic Cloud (10) (LMC). Using high-resolution spectra, we confirmed the discovery of a classical fundamental-mode Cepheid pulsator OGLE-LMC-CEP0227 in a well detached, double-lined, eclipsing system with near-perfect properties for deriving the masses of its two components with very high accuracy. (We obtained the spectra with the MIKE spectrograph at the 6.5-m Magellan Clay telescope at the Las Campanas Observatory in Chile, and with the HARPS spectrograph attached to the 3.6-m telescope of the European Southern Observatory on La Silla.) A finding chart for the system can be found on the OGLE Project webpage (10). Our spectroscopic and photometric observations of the binary system are best fitted by assuming a mass ratio of 1.00 for the two components (Fig. 1). This value was used to disentangle the pulsational and orbital radial-velocity variations of the Cepheid component of the binary. The resulting orbital radial-velocity curves of the components, and the pulsational radial-velocity curve of the Cepheid, are shown in Fig. 2. The spectroscopic and photometric observations were then analyzed using the 2007 version of the standard Wilson Devinney code (11,12). We accounted for the photometric variations of the Cepheid caused by the pulsations, as follows. First, we fitted a Fourier series of order 15 to the observations secured outside the eclipses. Second, we subtracted the corresponding variations in the eclipses in an iterative way, scaling the obtained fit according to the resulting Wilson–Devinney model. The I-band pulsational and orbital light curves, together with the best model obtained from the Wilson–Devinney code, are shown in Fig. 3. The corresponding astrophysical parameters of our system are presented in Table 1.

The mean radius of the primary (Cepheid) component that we obtained from our binary analysis shows excellent agreement with the radius predicted for its period from the Cepheid period–radius relation of ref. (13) (32.3 solar radii), strengthening our confidence in our results. In order to assign realistic errors to the derived parameters of our system, we performed Monte Carlo simulations. Our analysis of the very accurate existing data sets for OGLE-LMC-CEP0227 has resulted in a purely empirical determination of the dynamical mass of a classical Cepheid variable, with an unprecedented accuracy of 1%. We note that an end-to-end simultaneous solution for all parameters might reveal slightly different uncertainties, and would also illuminate the correlations in the uncertainties between the various derived quantities. From an evolutionary point of view, we have captured our system in a very short-lasting evolutionary phase, when both components are burning helium in their cores during their return from their first crossing of the Cepheid instability strip in the Hertzsprung–Russell diagram. The secondary component is slightly more evolved (it is larger and cooler), and is located just outside the Cepheid instability strip, so it is non-variable.

It is very important to note that OGLE-LMC-CEP0227 is a classical, high-mass Cepheid, and not a low-mass type-II Cepheid. This is clearly indicated by both its mass (Table 1) and its position on the period– luminosity diagram for OGLE Cepheids shown in Fig. 4 (which furthermore suggests that the star is a fundamental mode pulsator). Fundamental mode pulsation is also suggested by the strongly asymmetrical shapes and large amplitudes of the pulsation radial-velocity curve and of the I-band light curve (Figs 2 and 3). Of the three candidates for Cepheids in eclipsing binary systems detected earlier by the MACHO and OGLE projects (8,9), the objects MACHO-78.6338.24 and MACHO-6.6454.5 are type-II (low-mass) Cepheids[8,10]; only the object

OGLE-LMC_SC16-119952 (MACHO-81.8997.87) still appears to be a candidate for a classical Cepheid pulsating in the first overtone (14). However, there are currently several problems with the correct interpretation of this last object (9,14), and clearly more photometric and spectroscopic data are needed in order to reveal the true nature of this interesting system and eventually use it for a mass determination for a first overtone classical Cepheid. We also note that the type-II Cepheid MACHO-6.6454.5 belongs to the class of peculiar W Virginis stars introduced in ref. 15.

To estimate the pulsation mass of the Cepheid in LMC-OGLE- CEP0227, we adopted a period–mass relation based on nonlinear, convective Cepheid models constructed for the typical chemical com- position of LMC Cepheids (metallicity Z = 0.008, helium mass fraction Y = 0.256) (refs 5, 16, 17). This yields a pulsation mass of $M_p = 3.98 \pm 0.29$ solar masses for the star, which is independent of the assumed reddening and distance of the Cepheid and agrees within 1 σ with its dynamical mass, providing strong evidence that the pulsation mass of a Cepheid variable is indeed correctly measuring its true, current mass. This result contributes significantly to settling the controversy about classical Cepheid masses.

The overestimation of Cepheid masses by stellar evolution theory may be the consequence of significant mass loss suffered by Cepheids during the pulsation phase of their lives—such loss could occur through radial motions and shocks in the atmosphere (18,19). The existence of mild internal core mixing in the main-sequence progenitor of the Cepheid, which would tend to decrease its evolutionary mass estimate, is another possible way to reconcile the evolutionary mass of Cepheids with their pulsation mass (18).

**Acknowledgements** We gratefully acknowledge financial support for this work from the Chilean Center for Astrophysics FONDAP, the BASAL Centro de Astrofisica y Tecnologias Afines (CATA), NSF, Polish Ministry of Science, and the Foundation for Polish Science. The OGLE project has received funding from the European Research Council. It is a pleasure to thank the staff astronomers at Las Campanas and ESO La Silla who provided expert support in the data acquisition. We also thank Didier Queloz, Stefan Udry, and Chistophe Lovis for their help in reducing and analyzing the radial velocity data obtained with the HARPS instrument.

**Author Information**

Correspondence and requests for materials should be addressed to pietrzyn@astrouw.edu.pl

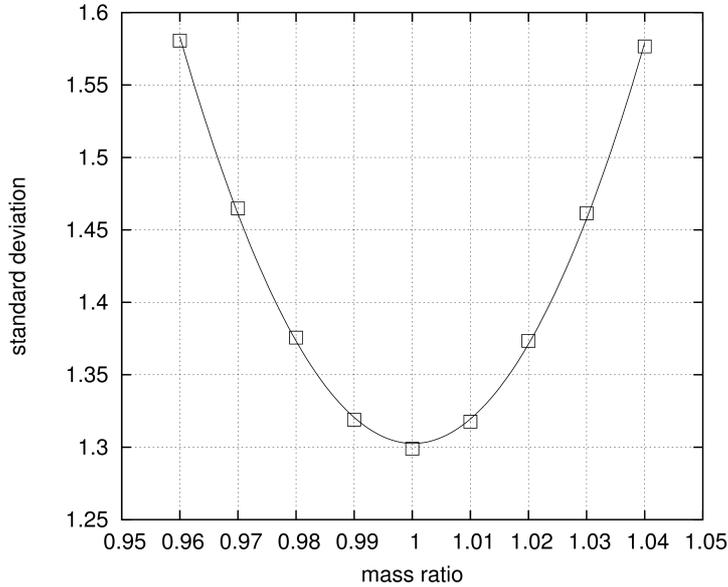

**Figure 1: The procedure adopted to separate pulsational and orbital motion of the Cepheid.** The following final ephemeris for our system was derived from the OGLE photometric data: orbital period P_orb = 309.673 ± 0.030 days, time of the minimum light of the binary system T0_orb = 2,454,895.91 ± 0.05 days; pulsational period P_pul = 3.797086 ± 0.000011 days, time of the Cepheid maximum light T0,pul = 2,454,439.94 ± 0.02 days. Adopting the photometric ephemeris, and having radial velocities measured for the secondary, non- pulsating component, we can scale them with the mass ratio and subtract them from the observed radial velocities of the Cepheid component, producing the pulsation radial-velocity curve of the Cepheid. Since both photometric and spectroscopic data indicate that the mass ratio of our system must be very close to 1, a set of pulsational radial-velocity curves of the Cepheid were obtained in this way for a range of mass ratios around 1, and the dispersion on each of these curves was measured. The resulting function of dispersion (expressed as standard deviation) versus mass ratio displayed in the figure shows a very well defined minimum around a mass ratio of 1.00. Independently, a mass ratio of our system of 0.99 ± 0.01 was derived from a least squares fitting of the orbit (systemic velocity, velocity amplitudes, eccentricity, periastron passage, and mass ratio) plus a Fourier series of order eight fitted to the pulsational radial- velocity variations of the Cepheid. We therefore adopted a mass ratio of 1.00 to disentangle the pulsational and orbital radial-velocity variations of the Cepheid component in the binary system.

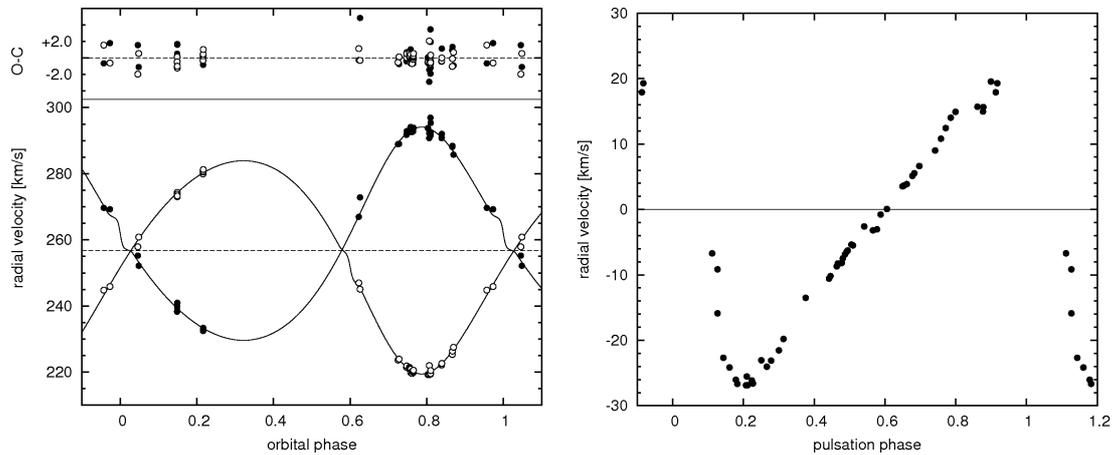

**Figure 2: Orbital motion of the two binary components, and the pulsational motion of the Cepheid variable in the binary system**  Left panel: the computed orbital radial-velocity curves of the two components of the LMC-OGLE-CEP227 binary system, after accounting for the intrinsic variation of the Cepheid's radial velocity due to its pulsation, together with the observed data. Filled and open circles, primary and secondary component, respectively. Top, the residuals of the observed velocities (O) from the computed ones (C), expressed in km/s. Right panel: the pulsational radial-velocity curve of the Cepheid in the binary system from 54 individual observations. The radial-velocity amplitude of 47 km/s  is typical for a 4-day fundamental mode classical Cepheid. All individual radial velocities were determined by the cross-correlation method using appropriate template spectra and the HARPS and MIKE spectra, yielding in all cases velocity accuracies better than 150 m/s (error bars are smaller than the circles in the figure).

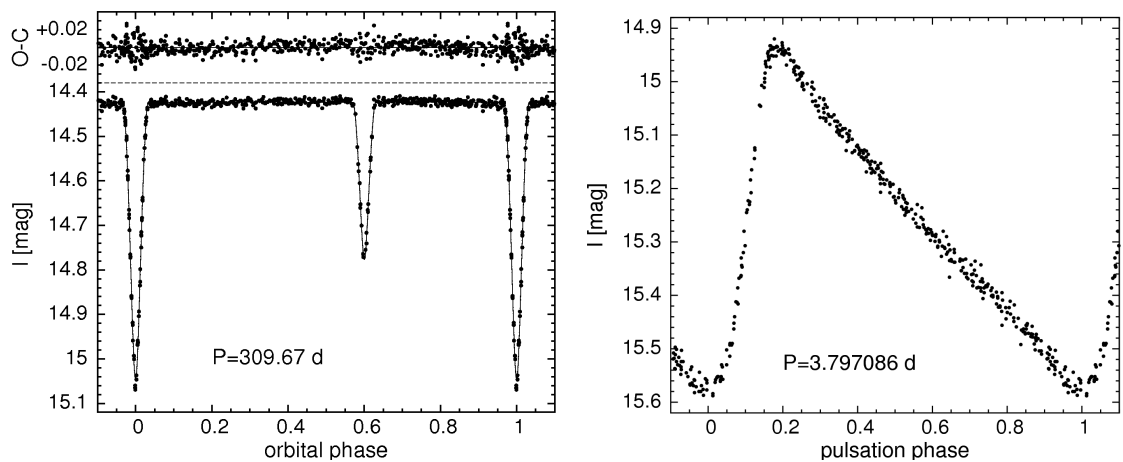

**Figure 3: Change of brightness of the binary system caused by the mutual eclipses, and the intrinsic change of the brightness of the Cepheid component caused by its pulsations.** Left panel: the orbital I-band light curve (367 epochs collected over 6.5 years) of the Cepheid-containing binary system LMC-OGLE-CEP0227, after removal of the intrinsic brightness variation of the Cepheid component together with the solution, as obtained with the Wilson–Devinney code. Top, the residuals of the observed magnitudes (O) from the computed orbital light curve (C). Right panel: the pulsational I-band light curve of the Cepheid in the binary system, folded on a pulsation period of 3.797086 days. The asymmetric, large-amplitude light curve is characteristic of a classical fundamental mode Cepheid pulsator.

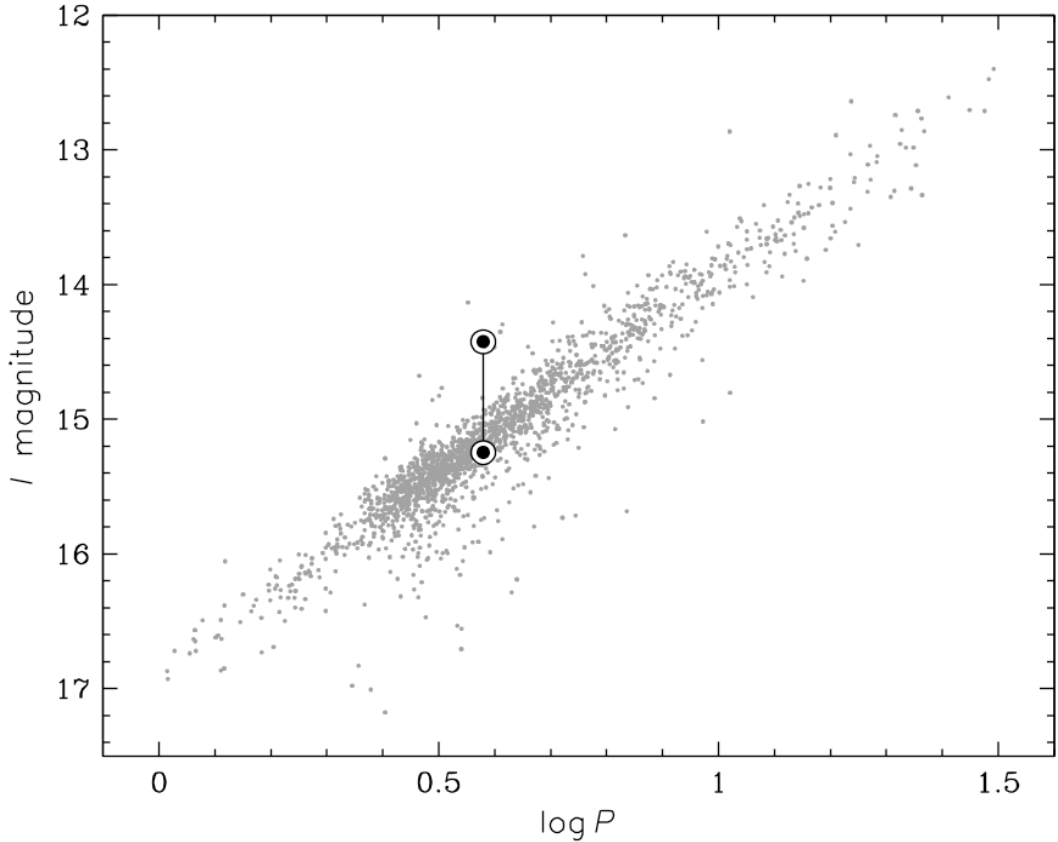

**Figure 4: The period and brightness of the Cepheid component of our system confirm that it is a classical Cepheid.** The period–luminosity relation (period in days) defined by the OGLE Project for fundamental mode classical Cepheids in the LMC10 in the photometric I band together with the position of OGLE-LMC-CEP0227. The upper circle corresponds to the total mean out-of- eclipse brightness of the system which contains the contribution of the binary companion to the Cepheid, while the lower circle measures the mean intensity magnitude of the Cepheid freed from the companion contribution. The Cepheid in the binary system fits well on the fundamental mode sequence, and is beyond any doubt a classical (and not type-II) Cepheid.

### Astrophysical parameters of the OGLE-LMC-CEP0227 system

| Parameter | Primary | Secondary |
|---|---|---|
| $M/M_\odot$ = mass | 4.14 ± 0.05 | 4.14 ± 0.07 |
| $R/R_\odot$ = *radius* | 32.4 ± 1.5 | 44.9 ± 1.5 |
| $T$ = *effective temperature* | 5900 ± 250 K | 5080 ± 270 K |
| e = excentricity | 0.1666 ± 0.0014 | |
| ω = periastron passage | 341.3 ± 1.1 deg | |
| γ = systemic velocity | 256.7 ± 0.1 km/s | |
| $P_{ORB}$  $P_{PUL}$ = *periods* | 309.673 ± 0.03 days   3.797086 ± 0.000011 days | |
| $i$ = inclination | 87.25 ± 0.25 deg | |
| $a/R_\odot$ = orbit size | 389.4 ± 1.2 | |
| $q$ = *mass ratio* | 1.00 ± 0.01 | |